\documentclass[overcite,twocolumn,showpacs,aps,prl,amsmath]{revtex4}
\usepackage{epsfig}

\begin{document}

\title{ Scaling Dark Energy}
\author{Salvatore Capozziello$^1$, Alessandro Melchiorri$^2$ and Alice Schirone$^1$}
\date{\today}


\affiliation{$^1$ Dipartimento di Fisica ``E.R. Caianiello'', Universit\`a di Salerno and
INFN, Sez. di Napoli, Gruppo Coll. di Salerno, via S. Allende, 84081 - Baronissi (Salerno), Italy\\
$^2$Dipartimento di Fisica ``G. Marconi'', Universita' di Roma ``La Sapienza'' and INFN, sez. di
Roma, Ple Aldo Moro 5, 00185, Roma, Italy}

\begin{abstract}

We investigate the possibility that dark energy is
scaling with epochs. A phenomenological model is introduced whose energy density depends 
on the redshift $z$ in such a way that a smooth transition among the three dominant phases 
of the universe evolution (radiation era, matter domination, asymptotic de Sitter state) 
is achieved. We use the WMAP cosmic microwave background data and the luminosity
distances of Type Ia Supernovae to test whether the model is in agreement with 
astrophysical observations.

\end{abstract}

\maketitle


\section{Introduction}

One of the greatest challenge in modern cosmology is to identify
the nature of the dark energy component which is causing the
observed accelerated expansion of the universe. Since a
cosmological constant, while in agreement with current
observations, is theoretically flawed, several alternatives have
been proposed. Some of the popular candidates to explain the
observations are a slowly-rolling scalar field, ``quintessence''
~\cite{Wetterich:fm}-\cite{Caldwell:1997ii}, or  a ``k-essence''
scalar field with non-canonical kinetic terms in the Lagrangian
~\cite{Armendariz-Picon:1999rj}-\cite{Chiba:1999ka} or 
``coupled quintessence'' where the scalar field is non-minimally
coupled with gravity ~\cite{luca}. 
The cosmological acceleration can be also achieved considering
geometrical counterparts in the gravitational Lagrangian other
than the standard Ricci scalar of General Relativity. This fact
allows to define a sort of effective curvature pressure and
curvature energy density which act as a time-varying cosmological
constant \cite{Capozziello:2002}.

One way to distinguish whether the dark energy is due to a
cosmological constant, to a scalar field  or something else is to
measure the equation of state, $w_X$, the ratio of the pressure
$p_X$ to the energy density $\rho_X$. A cosmological constant
always has $w_X=-1$ while scalar fields or curvature
counterparts generally have an equation of state which
differs from unity and varies with red shift $z$. Through
measurements of type Ia supernovae (SN-Ia), large-scale structure and
Cosmic Microwave Background (CMB) anisotropies, the
equation of state may be determined accurately enough in the next
few years to find out whether dark energy is actually different from
a cosmological constant or not.

An important general property of these so-called 'tracker' models
\cite{tracker} (and of any dark energy model which aims to
alleviate the fine tuning problem of the cosmological constant) is
that the scalar field equation of state (and its energy density)
remains close to that of the dominant background component during
most of the cosmological evolution.

For example, in power-law potential like $V=V_0 / \phi^{\alpha}$
the equation of state generally remains closer to the background
value $w_X=\alpha w_B/(\alpha+2)$ while the ratio of the energy
density of the scalar field to that of the dominant component
gradually increases. In models based on exponential potential
$V=V_0e^{-\lambda \phi}$ (\cite{attractor},\cite{napoli})  $w_X(z)$ mimics exactly
the scaling of the dominant background in the attractor regime
($w_X=w_B$) and if the background component scales as
$\rho_B=\rho_0({a_0 \over a})^n$, then the scaling field
approaches an attractor solution, and its fractional energy
density is given by $\Omega_X={n \over \lambda^2}$, constant with
redshift. As further example, in ``k-essence'' 
models, the \textit{k}-essence\ undergoes two transitions in its behavior,
one at the onset of matter-domination and a second when
\textit{k}-essence\ begins to dominate over the matter density. 
During the radiation-dominated era, the  $k$-essence energy tracks the
radiation, falling as $1/a^4$ where $a$ is the scale factor. 
The onset of the matter-dominated era  automatically triggers a change in the
behavior of \textit{k}-essence\ such that it begins to act as an
energy component with $w_X(z)  \le 0$. When \textit{k}-essence\ overtakes 
the matter density, $w_X(z)$ changes to another value around $-1$, 
the precise value of which depends on the detailed model.

Given the new release of cosmological data from high-precision
measurements of CMB anisotropies (see e.g.
\cite{map1}) and SN-Ia luminosity distances
(\cite{riess}) (which are now providing a $\sim 18 \sigma$
evidence for a dark energy component) is therefore extremely
timely to check if any hint for a ``{\it scaling}'' dark energy is
present in the data. 

Moreover, recent analysis of SN-Ia data
(see e.g. \cite{meta} and \cite{cooray}) with model independent 
parameterization have found that dark energy which evolves with time 
provides a better fit to the SN-Ia data than a 
standard cosmological constant.

In this paper we use a phenomenological approach to constrain a
dark energy component with an evolutionary behavior similar to
the models mentioned above. In particular, we use a toy model
whose energy density depends on the redshift $z$ in such a way
that a smooth transition among the main three cosmological scaling
regimes (radiation, matter and dark energy)
is achieved \cite{cardone}. We then use the Wilkinson Microwave 
Anisotropy Probe (WMAP) CMB data and the luminosity distances 
of SN-Ia to test whether the model is in agreement with astrophysical
observation and/or any evidence for ``scaling'' dark energy is
present in the data.

This approach has the main advantage of being theoretically well
motivated since a scaling model can reasonably approximate the 
behavior of most of the dark energy theories on the market 
(cosmological constant included). An analysis with an higher 
number of parameters to describe the dark energy evolution would 
probably be better suited to detect variations from a cosmological 
constant or to test models with a rapidly evolving equation of state. 
However, allowing more degrees of freedom could introduce serious 
degeneracies, fit unknown systematics and produce final results 
of difficult theoretical interpretation. The analysis and results 
presented here can be therefore considered as complementary to 
recent analysis which sampled a wider set of parameters 
(see e.g. \cite{corasaniti}) or on the contrary restricted 
the study to a constant with redshift equation of state
(see e.g. \cite{map1}, \cite{mmot}).
The plan of the paper is as follows: in the next section, the
phenomenological model is illustrated in detail. In Section $3$,
the analysis method is briefly discussed while in Section $4$ we
present  the results. Section $5$ is devoted to conclusions.

\section{A phenomenological model for scaling dark energy}

Let us now illustrate in  detail the phenomenological method used
for our analysis. 
The energy density of our scaling dark energy model evolves 
with redshift as (see \cite{cardone}):

\begin{equation}
\rho_X(z) = A \ \left ( 1 + \frac{1 + z}{1 + z_s} \right ) \ 
\left [ 1 + \left ( \frac{1 + z}{1 + z_b} \right )^{3} \right ]
\label{eq: rhoz}
\end{equation}

\noindent where $A$ is a normalization constant, related to the 
today dark energy density $\Omega_X$, and $z_s$ and $z_b$ are 
free $2$ parameters that identify the three epochs of scaling. 
The corresponding dark energy equation of state is indeed:

\begin{equation}
w_X(z) = \displaystyle{
\frac{\left [ \left ( \frac{1 + z}{1 + z_b} \right )^3 - 2 \right ] \frac{1 + z}{1 + z_s} - 3}
{3 \left ( 1 + \frac{1 + z}{1 + z_s} \right ) \left [ 1 + \left ( \frac{1 + z}{1 + z_b} \right )^3 
\right ]}}
\ .
\label{eq: wzmodel}
\end{equation}

\noindent which depends on the parameters $z_s$ and $z_b$ such that:

\begin{displaymath}
w_X \sim 1/3 \ \ {\rm for} \ \ z >> z_s \ ,
\end{displaymath}
\begin{displaymath}
w_X \sim 0 \ \ {\rm for} \ \ z_b << z << z_s \ ,
\end{displaymath}
\begin{displaymath}
w_X \sim -1 \ \ {\rm for} \ \ z << z_b \ .
\end{displaymath}

Therefore the model we obtain is able to mimic a
fluid following first a radiation equation of state,
then a matter phase and finally approaching a deSitter
phase with constant energy. 
This model is useful to identify and/or constrain
a cosmological imprint of scaling dark energy in the data.
The same model can be further extended to a fluid
with transitions between more generic equations of state.

\begin{figure}[t]
\begin{center}
\includegraphics[scale=0.50]{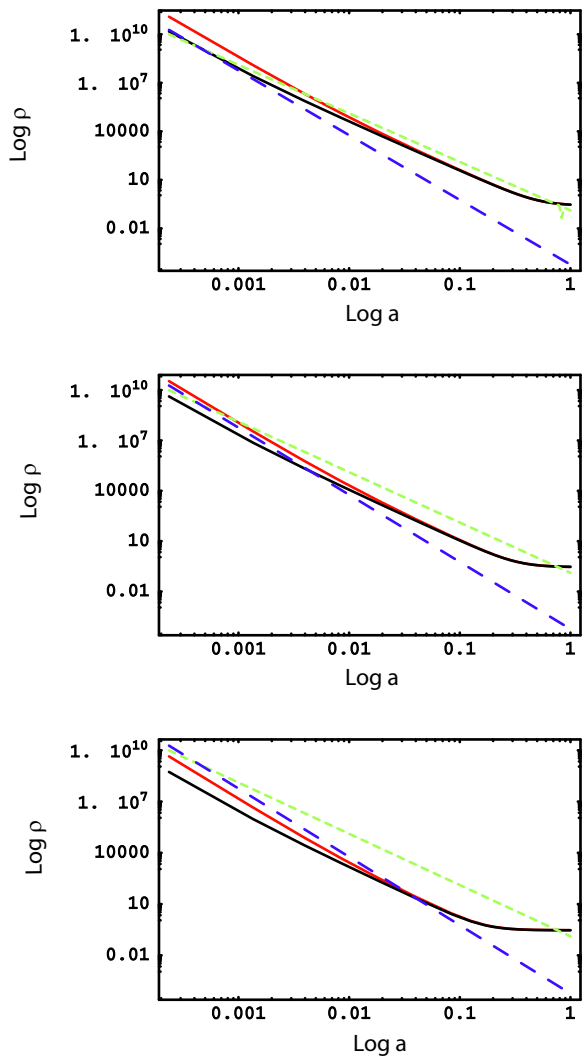}
\end{center}
\caption{Evolution of the overall energy density of the Universe
with redshift in different theoretical frameworks. In the panels,
the phenomenological dark energy model described in the text has
$z_s=100$ (grey line) and $z_s=1000$ (black line). The top, middle
and bottom panels show the cases with $z_b=1$, $z_b=2$ and $z_b=5$
respectively. In all plots, the short dashed line is the matter 
component while the long-dashed line is radiation (photons and 3 
massless neutrinos) contribution.} \label{figo0}
\end{figure}

In Fig. \ref{figo0},
the behavior of the energy density of the $3$ components in our
model (matter, radiation and dark energy) in function of the scale
factor are plotted. As one can see, if the matter to radiation
transition redshift is much smaller than the $\Lambda$-CDM
redshift of equivalence $z_s < z_{Eq}\sim 3200$, the dark energy
can be again dominant in the past. However this can be tuned by
$z_b$ which shifts the dark energy transition between a
cosmological constant and matter. For example, a radiation to
matter transition at redshift $z_s \sim 100$ can still be in
agreement with a negligible dark energy contribution in the past,
providing that $z_b > 5$.

The time variation in $w_X$ is small in comparison to the
expansion rate of the universe. We assume purely adiabatic
contributions to the perturbations in the spectrum. The sound speed 
is therefore fully determined by $w_X(z)$ and we 
integrate the evolution equations for the density and
velocity (we neglect shear) perturbations in the dark energy fluid as in
~\cite{Ma95}. While the adiabatic approximation is not 
completely correct for most scalar field or quintessence models, 
since there dissipative processes generate entropic perturbations 
in the fluid and a more general relation is needed, such a model 
can be considered anyway as a good start for analyzing the 
current data.

\begin{figure}[t]
\begin{center}
\includegraphics[scale=0.40]{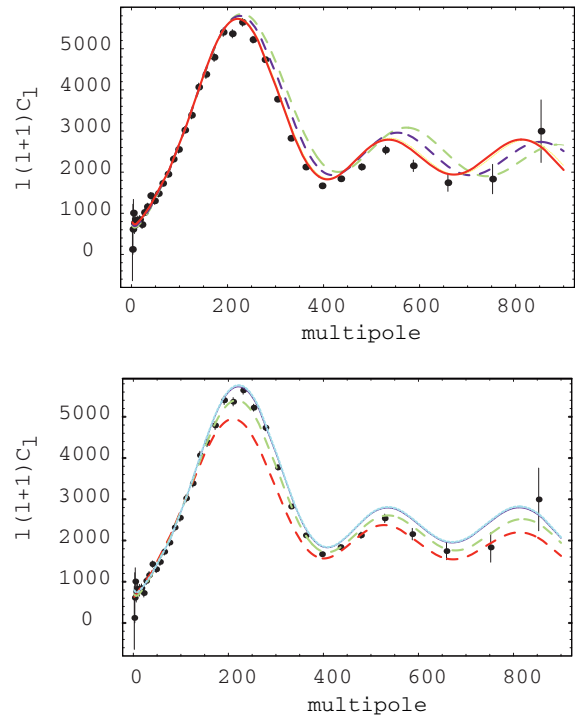}
\end{center}
\caption{Top Panel: CMB power spectra normalized at $\ell=111$ for
a universe with $\Omega_m=0.35$ and $\Omega_{X}=0.65$. The
transition redshift matter-radiation has been fixed to $z_s=5000$.
Models with different values of $z_b$ are plotted: red ($z_b=1$),
green ($z_b=2$), blue ($z_b=5$), light blue ($z_b=20$). Bottom
Panel: CMB power spectra normalized at $\ell=111$ for a universe
with $\Omega_m=0.35$ and $\Omega_{X}=0.65$. The  matter-dark
energy transition redshift has been fixed to $z_b=5$. Models with
different values of $z_s$ are plotted: green ($z_s=50$), blue
($z_s=100$), yellow ($z_s=500$), red ($z_s=5000$). The WMAP data
points are also plotted for comparison.} \label{figoa}
\end{figure}

In Fig. \ref{figoa}, top panel, we plot several power spectra
computed with CMBFAST \cite{CMBFAST} in function of $z_b$ with 
matter density $\Omega_m=0.35$, $\Omega_{X}=0.65$ and the 
dark energy matter-radiation transition redshift
fixed at $z_s=5000$, well after the 
redshift of equivalence in standard cold dark matter model dominated
by a cosmological constant ($\Lambda$-CDM).
As we can see, increasing $z_b$ has the effect of mimicking more
and more the cosmological constant behavior. On the other hand,
decreasing $z_b$ increases the effective equation of state
$w_{eff}$ shifting the peaks in the CMB spectrum toward smaller
angular scales (see e.g. \cite{rachel}).

In Fig. \ref{figoa}, bottom panel,   $z_b=5$ is fixed and  the
dependence of $z_s$ is studied. In this case, the effect is
smaller, and it is clear that as soon as $z_s$ becomes smaller
than the redshift of equivalence in $\Lambda$-CDM, $\Omega_{X}$
can dominate again as a relativistic component 
and leave an imprint on the CMB spectrum
through the Early Integrated Sachs-Wolfe effect 
(see e.g. \cite{bowen}).

\section{The Analysis Method}

In order to bound this phenomenological dark energy model, we
consider a template of flat, adiabatic, $X$-CDM models computed
with CMBFAST~\cite{CMBFAST}. We sample the relevant parameters as
follows: $\Omega_{cdm}h^2 = 0.05,...0.20$, in steps of  $0.01$;
$\Omega_{b}h^2 = 0.015, ...,0.030$ (motivated by Big Bang
Nucleosynthesis), in steps of  $0.001$ and $\Omega_{X}=0.05, ...,
0.95$, in steps of  $0.02$.

\noindent We sample the $2$ parameters of the dark energy model in
the range $z_b = 0.05,...,7.55$, in steps of  $0.5$ and $z_s = 20,
...,740$ in steps of $80$.

\noindent The value of the Hubble constant in our database is not
an independent parameter, since it is determined through the
flatness condition.  The conservative top-hat bound $0.55 < h <
0.85$ is adopted and the $1\sigma$ constraint on the Hubble
parameter, $h=0.71\pm0.07$, obtained from Hubble Space Telescope
(HST) measurements~\cite{freedman}, is also considered.

\noindent We allow for a reionization of the intergalactic medium by
varying the Compton optical depth parameter
$\tau_c$ over the range $\tau_c=0.05,...,0.25$ in steps of $0.02$.

\noindent For the CMB data, we take into account the recent
temperature and cross polarization results from the WMAP satellite
(\cite{map1}) using the method explained in (\cite{map5}) and the
publicly available code on the LAMBDA web site. 

\noindent  Finally  constraints obtained from the luminosity
measurements of Type Ia Supernovae (SN-Ia) are incorporated from
~\cite{riess} using the GOLD dataset. The SN-Ia luminosity data
are independent of the neutrino energy density but, as the CMB
data, are helpful in breaking degeneracies between the parameters
we are going to consider.

\section{The Results}

\begin{figure}[t]
\begin{center}
\includegraphics[scale=0.35]{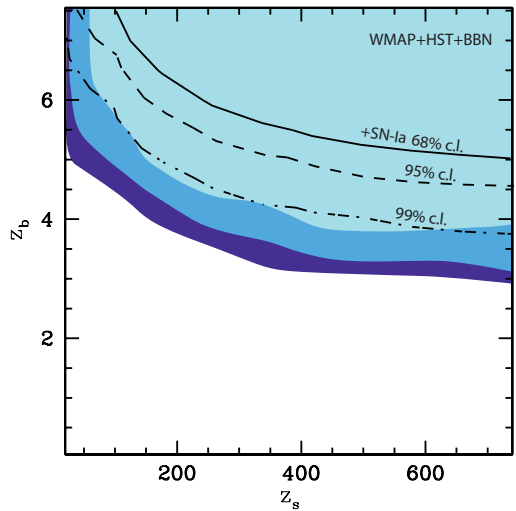}
\end{center}
\caption{Likelihood probability contours at $68 \%$ (light blue), $95 \%$ 
(blue) and $99 \%$ (dark blue) in the $z_s$-$z_b$ plane from WMAP.
The solid,long
dashed and short-dashed lines are the $68 \%$,  $95 \%$ and $99 \%$
from a WMAP+SN-Ia analysis. }
\label{figo1}
\end{figure}

\noindent In Figure~\ref{figo1},  the likelihood contours are
plotted in the ($z_s$, $z_b$) plane only for the CMB data. 
As we can see the current data doesn't show any evidence for 
scaling dark energy and we can only derive weak lower limits
on the two parameters of the model. We have found that $z_s> 60$ and 
$z_b > 3.8$ at $1- \sigma$. 
From the same Figure is clear that an interesting correlation 
exists between the  $2$ parameters. Namely, the current CMB data 
do not favor or provide evidence for an extra matter or radiation 
component: if $z_s$ is too small then it is necessary to consider 
larger values of $z_b$ in order to have the extra 
dark energy component not dominant in the past. 
At high $z_s$, we have an asymptotic value  $95 \%$ c.l. 
of $z_b > 3.0$.
Also plotted on the figure are the likelihood contours derived
by a combined CMB+SN-Ia analysis. The inclusion of the SN-Ia data 
improves the constraints to $z_b > 5$ and $z_s >100$ at 
$1- \sigma$. SN-Ia data are insensitive to variations in
$z_s$ since they are probing only redshifts $z <1.5$,
but provide anyway complementary constraints
on $z_b$ and the matter density $\Omega_m$. 

\begin{figure}[t]
\begin{center}
\includegraphics[scale=0.35]{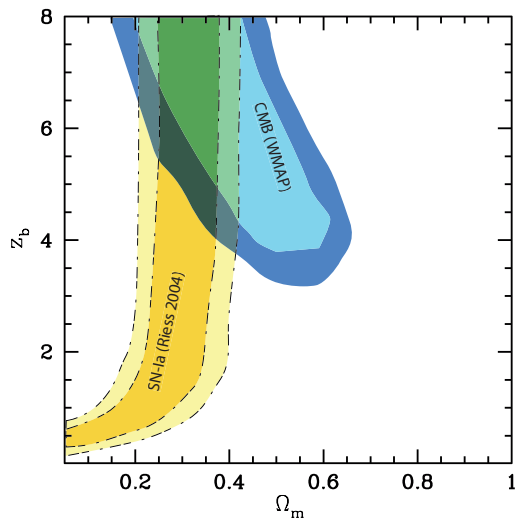}
\end{center}
\caption{$68 \%$ and $95 \%$ likelihood probability contours in the 
$\Omega_m$-$z_b$ plane from WMAP and SN-Ia.}
\label{figo2}
\end{figure}

\noindent This is better explained in  Figure~\ref{figo2}, 
where we over-impose the likelihood contours
in the ($\Omega_m$, $z_b$) plane from CMB and SN-Ia analysis.
As we can see, the current SN-Ia
data does not provide evidence for dark energy evolution. 
However, low values of $z_b$ are compatible with the SN-Ia data if one 
decreases the amount of the matter component. 
The SN-Ia data is consistent with $\Omega_m=0$ but in this 
case the dark energy model behaves like an unified dark energy 
model with transition redshift $z_b \sim 0.5$.
On the other hand, as we can see from the plot, 
lower values of $z_b$ are compatible with
CMB data if one increases the matter density. This
is easily explained from the fact that a lower $z_b$
results in higher values for the effective dark energy 
equation of state. The direction of degeneracy in
the plot in the case of the CMB data is therefore only
a consequence of the geometrical degeneracy present
in angular diameter distance data at high redshift.
The lower limit on $z_b$ from the CMB data 
comes mainly from our assumptions on the possible
values of the Hubble parameter.
Combining the CMB and SN-Ia further breaks this 
degeneracy, improving the lower limit on $z_b$
and excluding at high significance an unified dark energy
model with $\Omega_m=0$.

\section{Conclusions}

A phenomenological {\it scaling dark energy} model is discussed in
this paper matching it with the current CMB and SN-Ia data to
identify the signatures of a possible cosmological evolution of
dark energy, as expected and predicted in several theories. We do
not take into account particular scalar field or quintessence
models but discuss how the state equation of cosmic fluid,
depending on redshift, scales with respect to the epoch passing
from a radiation regime to a dark-energy-matter-dominated era. The
approach is extremely general since the dynamical behavior which
we discuss should be the one expected for the most of the models
present in literature.

We found that the current data does not show evidence for
cosmological evolution of dark energy, providing the $68 \%$c.l.
bounds $z_b >5.0$ and $z_s>100$  constraining the 
presence of scaling dark energy in the universe. A simple but 
theoretically flawed cosmological constant still provides a 
good fit to the data (see also the discussion in \cite{napoli}). 

Interesting correlations between the parameters used in the analysis of single 
datasets are present.For example, lower values of the matter 
density ($\Omega_m \sim0.1$) could hint for an evolution  
in the SN-Ia data, but these values are ruled out by CMB measurements.

Dark energy models with a subdominant contribution to the overall energy 
density of the universe for most of the cosmological
evolution are clearly preferred. While this condition is easy 
to achieve for many models based on scalar fields or 
topological defects (see e.g. \cite{conversi}), 
the quantity of information we can hope
to extract from future data about dark energy is more limited.
 
However, our approach could be further improved and extended
to more general scaling solutions. In this respect, the 
incoming wide cosmological surveys as PLANCK or SNAP will 
provide essential data.

\begin{acknowledgments}
{\it \bf Acknowledgments} 
We wish to thank Luca Amendola, Vincenzo Cardone and Antonio Troisi for 
useful discussions.
\end{acknowledgments}


\begin{thebibliography}{99}


\bibitem{tracker}
I.~Zlatev, L.~Wang and P.J.~Steinhardt, Phys. Rev. Lett. {\bf 82}, 896 (1999);
Phys. Rev. D {\bf 59}, 123504 (1999).

\bibitem{attractor}
J.~J.~Halliwell, Phys. Lett. {\bf B185} 341 (1987);
J.~Barrow, Phys. Lett. {\bf B187} 12 (1987);
B.~Ratra, P.~Peebles, Phys. Rev. D {\bf 37} 3406 (1988);
C.~Wetterich, Astron. \& Astrophys. {\bf 301} 321 (1995).

\bibitem{luca}
L.~Amendola,
Phys.\ Rev.\ D {\bf 62} (2000) 043511
[arXiv:astro-ph/9908023].

\bibitem{napoli}
 C. Rubano, P. Scudellaro, E.
Piedipalumbo, S. Capozziello, M. Capone, Phys. Rev. D {\bf  69},
103510 (2004).


\bibitem{Wetterich:fm}
C.~Wetterich,
Nucl.\ Phys.\ B {\bf 302}, 668 (1988).

\bibitem{Caldwell:1997ii}
R.~R.~Caldwell, R.~Dave and P.~J.~Steinhardt,
Phys.\ Rev.\ Lett.\  {\bf 80}, 1582 (1998)
[arXiv:astro-ph/9708069].

\bibitem{Armendariz-Picon:1999rj}
C.~Armendariz-Picon, T.~Damour and V.~Mukhanov,
Phys.\ Lett.\ B {\bf 458}, 209 (1999)
[arXiv:hep-th/9904075].

\bibitem{Chiba:1999ka}
T.~Chiba, T.~Okabe and M.~Yamaguchi,
Phys.\ Rev.\ D {\bf 62}, 023511 (2000)
[arXiv:astro-ph/9912463].

\bibitem{Capozziello:2002}
S. Capozziello, Int. Jou. Mod. Phys. D {\bf 11}, 483 (2002); S.
Capozziello, V.F. Cardone, S. Carloni and A. Troisi, Int. Jou.
Mod. Phys. D {\bf 12}, 1969 (2003); S.M. Carroll, V. Duvvuri, M.
Trodden and M.S. Turner, (2003) astro-ph/0306438; S. Nojiri and
S.D. Odintsov, Phys. Rev. D {\bf 68}, 123512 (2003);  E.E.
Flanagan, Phys. Rev. Lett.  {\bf 92}, 071101 (2004); G. Allemandi,
A. Borowiec and M. Francaviglia, (2004) hep-th/0403264.

\bibitem{cardone}
V.~F.~Cardone, A.~Troisi and S.~Capozziello,
Phys.\ Rev.\ D {\bf 69} (2004) 083517
[arXiv:astro-ph/0402228].


\bibitem{meta}U.~Alam, V.~Sahni and A.~A.~Starobinsky,
JCAP {\bf 0406} (2004) 008
[arXiv:astro-ph/0403687].

\bibitem{cooray}
D.~Huterer and A.~Cooray,
arXiv:astro-ph/0404062.

\bibitem{corasaniti}
P.~S.~Corasaniti, M.~Kunz, D.~Parkinson, E.~J.~Copeland and B.~A.~Bassett,
arXiv:astro-ph/0406608.


\bibitem{mmot}
A.~Melchiorri, L.~Mersini, C.~J.~Odman and M.~Trodden,
Phys.\ Rev.\ D {\bf 68} (2003) 043509
[arXiv:astro-ph/0211522];

\bibitem{rachel}
R.~Bean and A.~Melchiorri,
Phys.\ Rev.\ D {\bf 65}, 041302 (2002)
[arXiv:astro-ph/0110472].

\bibitem{bowen}
R.~Bowen, {\it et al}.,
Mon.\ Not.\ Roy.\ Astron.\ Soc.\  {\bf 334} (2002) 760
[arXiv:astro-ph/0110636].


\bibitem{Ma95} C.P Ma, E. Bertschinger \apj {\bf 455} 7 (1995).


\bibitem{CMBFAST} U.~Seljak and M.~Zaldarriaga, \apj {\bf 469}, 437 (1996).

\bibitem{freedman}
W. Freedman {\it et al.},
Astrophysical Journal, 553, 2001, 47.

\bibitem{2dFGRS}J.~Peacock {\it et al.}, Nature {\bf 410}, 169 (2001).

\bibitem{thx}M.~Tegmark, A.~J.~S.~Hamilton and Y.~Xu,
arXiv:astro-ph/0111575

\bibitem{sloan}
M.~Tegmark {\it et al.}  [SDSS Collaboration],
Phys.\ Rev.\ D {\bf 69} (2004) 103501
[arXiv:astro-ph/0310723].

\bibitem{map1}C.~L.~Bennett {\it et al.}, astro-ph/0302207

\bibitem{map5}L.~Verde, {\it et al.}, astro-ph/0302218

\bibitem{riess}
A.~G.~Riess {\it et al.}  [Supernova Search Team Collaboration],
arXiv:astro-ph/0402512.

\bibitem{conversi} 
L.~Conversi, A.~Melchiorri, L.~Mersini and J.~Silk,
Astropart.\ Phys.\  {\bf 21} (2004) 443
[arXiv:astro-ph/0402529];


\end{thebibliography}
\end{document}